\documentclass[10pt,journal]{IEEEtran}

\usepackage{subfigure}
\usepackage{amsfonts}
\usepackage{amssymb}
\usepackage{amsmath}
\usepackage{calrsfs}
\usepackage{graphicx}
\usepackage{epstopdf}
\usepackage{bm}
\usepackage{makecell}
\usepackage{xcolor}
\usepackage{flushend}

\usepackage{algorithm,algpseudocode}
\algnewcommand{\Inputs}[1]{%
  \State \textbf{Inputs:}
  \Statex \hspace*{\algorithmicindent}\parbox[t]{.8\linewidth}{\raggedright #1}
}
\algnewcommand{\Initialize}[1]{%
  \State \textbf{Initialize:}
  \Statex \hspace*{\algorithmicindent}\parbox[t]{.8\linewidth}{\raggedright #1}
}

\algdef{SE}[PROCEDURE]{Procedure}{EndProcedure}%
   [2]{\algorithmicprocedure\ \textproc{#1}\ifthenelse{\equal{#2}{}}{}{(#2)}}%
   {\algorithmicend\ \algorithmicprocedure}%

\AtEndDocument{\par\leavevmode}

\begin{document}

\onecolumn

\title{Reliable and Low-Latency Fronthaul \\for Tactile Internet Applications}

 \author{\IEEEauthorblockN{Ghizlane Mountaser,  Toktam Mahmoodi, and Osvaldo Simeone\\}
	\IEEEauthorblockA{Centre for Telecommunications Research, Department of Informatics \\King's College London, London WC2B 4BG, UK}}


\maketitle

\begin{abstract}
With the emergence of Cloud-RAN as one of the dominant architectural solutions for next-generation mobile networks, the reliability and latency on the fronthaul (FH) segment become critical performance metrics for applications such as the Tactile Internet. Ensuring FH performance is further complicated by the switch from point-to-point dedicated FH links to packet-based multi-hop FH networks. This change is largely justified by the fact that packet-based fronthauling allows the deployment of FH networks on the existing Ethernet infrastructure. This paper proposes to improve reliability and latency of packet-based fronthauling by means of multi-path diversity and erasure coding of the MAC frames transported by the FH network. Under a probabilistic model that assumes a single service, the average latency required to obtain reliable FH transport and the reliability-latency trade-off are first investigated. The analytical results are then validated and complemented by a numerical study that accounts for the coexistence of enhanced Mobile BroadBand (eMBB) and Ultra-Reliable Low-Latency (URLLC) services in 5G networks by comparing orthogonal and non-orthogonal sharing of FH resources.
\end{abstract}


\section{Introduction}

One of the major paradigm shifts in the mobile and wireless networking is the transition from a static network configuration to a flexible and reconfigurable network set-up through softwarization, virtualization, and cloudification \cite{Mahmoodi2014OnUA, 7996513, bookcran}. Among the solutions that made their way into the architectural choices for the fifth generation (5G) of wireless networks, is the cloudification of the Radio Access Network (RAN) through the Cloud-RAN architecture \cite{3gpp.38.801}. Cloud-RAN offers the flexibility to move the baseband unit (BBU) functionalities to a central unit (CU) in support of multiple remote radio heads (RRHs) or Radio Units (RUs). The RUs are connected to the CU via an interface called fronthaul (FH). While enabling centralized processing and control, the introduction of the FH introduces an additional segment in the network, over which the main Key Performance Indicators (KPIs) should be delivered \cite{7456186}. In particular, one of the key challenges in 5G is achieving high reliability and low latency KPIs at the same time in order to enable innovative applications and use cases, such as the Tactile Internet \cite{7403840, 8399482}, industrial networking \cite{10.1002/ett.3057}, or smart cities \cite{8038764, cond2015}.

The 3GPP standardisation body has provided guidelines for delivering  Ultra-Reliable Low Latency Communications (URLLC) \cite{38300, 8329620}. The new specification defines the New Radio interface, which introduces structural change in the physical layer (PHY) in terms of numerology in order to meet stringent constraints on latency and reliability. However, there is limited research on how FH technologies can support the levels of latency and reliability needed for 5G URLLC services. This paper aims at addressing this knowledge gap.

\begin{figure}[t]
	\begin{centering}
		\includegraphics[width=0.4 \textwidth]{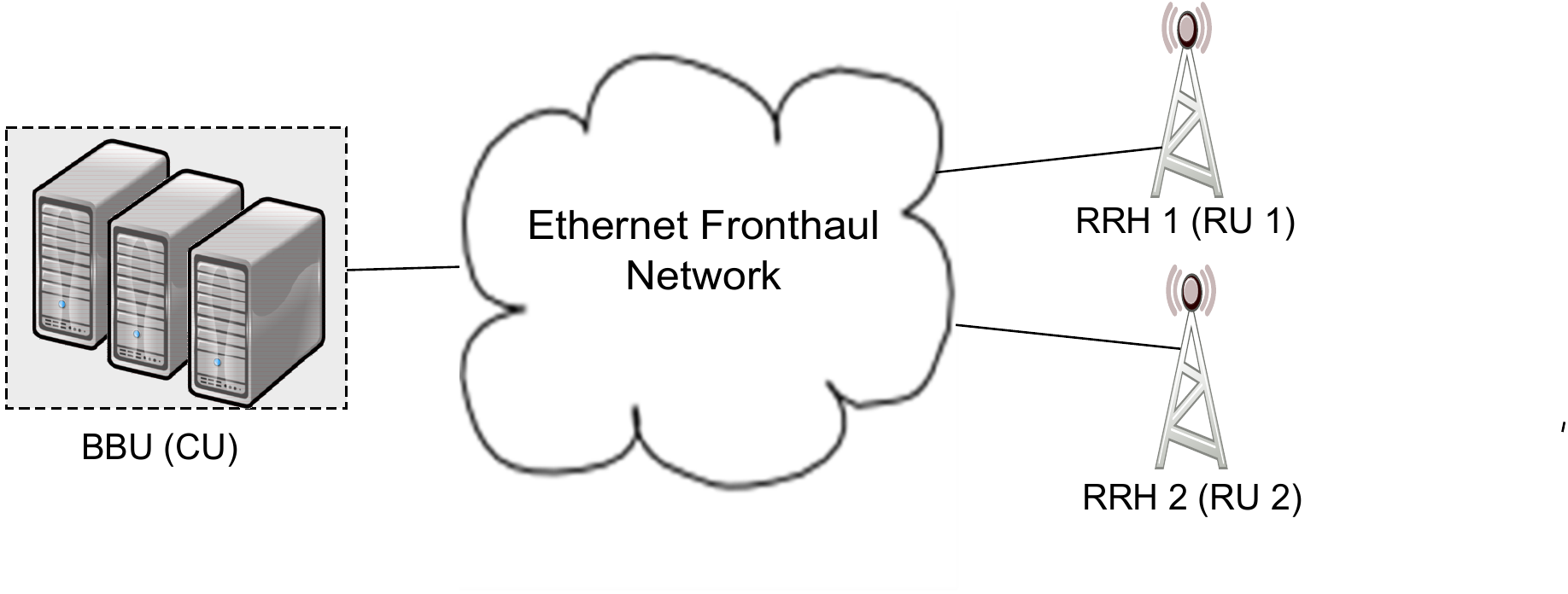}
		\caption{Cloud-RAN architecture with Ethernet-based multi-path FH.}   \label{fig:multi_hop_switch_fh}
	\end{centering}
\end{figure}

The conventional FH topology consists of dedicated lines from the CU to each RU, i.e., of point-to-point links, that transports baseband radio samples in a serial manner. A more economic solution has recently emerged, whereby dedicated lines are replaced by a multi-path packet-based FH network that can leverage the wide deployment of the Ethernet infrastructure \cite{MEF, kar50278} (see Fig. \ref{fig:multi_hop_switch_fh}). Multi-hop packet-based networks make it ever more challenging to ensure the high reliability and low-latency KPIs expected by 5G systems.

Increasing reliability is typically accomplished through retransmissions based on feedback or through redundancy. Depending on the split point between CU and RU, the latency requirements on the FH link differ \cite{3gppTsgRan3}, but the figures are in the range between 55 $\mu$s and 1 ms \cite{8269135}. Given this strict delay requirement, retransmission based on feedback is not a viable choice. In contrast, redundancy through transmission over multiple FH links offers a feasible solution. In a multi-hop packet-based network, this can be realized by transmitting over multiple paths, or routes, between the CU and an RU.

While different options of split point between CU and RU have been introduced in the 3GPP standard \cite{3gpp.38.801}, our focus here is on the split between MAC and PHY. Accordingly, the FH transports MAC frames. Two of the authors of this work have previously studied feasibility of the Ethernet-based FH for this configuration in \cite{7925770} by using a hardware-based experimentation.

In this context, as a first option to leverage multi-path transmission, one could replicate the same MAC data stream across multiple FH paths, yielding \emph{Multi-Path Transport with Duplication} (MPD). MPD ensures correct reception as long as any path succeeds in delivering the FH data. This increases reliability, while also increasing FH network congestion and hence potentially affecting the latency KPI. As a dual solution, one could split each FH MAC frame in disjoint blocks transmitted on different paths. Since each FH path would need to carry less information, this approach would generally reduce congestion and transport latency on each path, but reliable transport would rely on the correct reception on all paths.

As a means to bridge the gap between the two extreme solutions highlighted above, in this paper we propose to use coding techniques on the MAC frames transported by the FH network. Coding can reduce the FH transport overhead as compared to MPD, while still providing resilience to the potential unreliability on some of the FH paths. We specifically adopt erasure coding methods, such as rateless or fountain, coding. Erasure coding is able to recover from a number of packet losses equal to the amount of redundancy introduced by coding. Fountain codes are adopted by standards such as 3GPP Multimedia Broadcast Multicast Service~(MBMS) \cite{3gpp-mbms} for broadcast file delivery and streaming services, and by the IETF RFCs \cite{rfc-5053}. They have been used in data storage applications \cite{6483236}, content distribution networks, collaborative downloading, and for exploiting multiple interfaces such as Multipath Transmission Control Protocol (MTCP) \cite{6729115}. We refer to the considered approach as \emph{Multi Path Transport with Coding (MPC)}.

To elaborate, in this work, we consider a Cloud-RAN model with multiple packet-based FH paths between the CU and the RUs as illustrated in Fig. 1. We analyze the performance of baseline single-path (SP) transmission (Fig. 2), MPD (Fig. 3) and MPC (Fig. 4) for downlink communication by first considering a standard set-up with a single service. Based on the insights from the analysis, we then present extensive experimental results concerning the performance of Cloud-RAN with multi-path FH under MPD and MPC in the presence of both enhanced Mobile BroadBand (eMBB) and URLLC services as described by the 3GPP standard. The study compares orthogonal and non-orthogonal sharing of FH resources. The closest research works to this paper are a study presented in \cite{8292238}, where a combination of diversity and network coding is used for improving throughput on the FH link; and reference \cite {8006715}, which discusses how coded FH transmission together with caching can reduce latency.

The remainder of this paper is organized as follows. Section \ref{state_of_the_art} provides an overview of different transport technologies and related work on Cloud-RAN split solutions. Section \ref{system_model} elaborates on the different FH solutions, namely SP, MPD, and MPC. Section \ref{performance_metrics} describes analysis of the single-service case in terms of average latency and reliability. Numerical results are detailed in Section \ref{analysis} by considering the coexistence of eMBB and URLLC traffic types. Finally, conclusive remarks are discussed in Section \ref{conclusion}.


\section{Background} \label{state_of_the_art}

Cloud-RAN is one of the key elements of the 5G architecture, owing to its capability to enhance spectral efficiency and cell-edge performance, as well as to facilitate operation in the presence of mobility, due to the coordination among adjacent cells. However, Cloud-RAN relies on the availability of a FH infrastructure to connect RUs to the CU. Therefore, supporting URLLC services in a C-RAN architecture imposes stringent requirement on both latency and reliability on the FH. In this section, we overview different FH technologies and discuss some of their advantages and disadvantages as enablers of URLLC services on C-RAN architectures.

\begin{figure}[t]
	\begin{centering}
		\includegraphics[width=0.4 \textwidth]{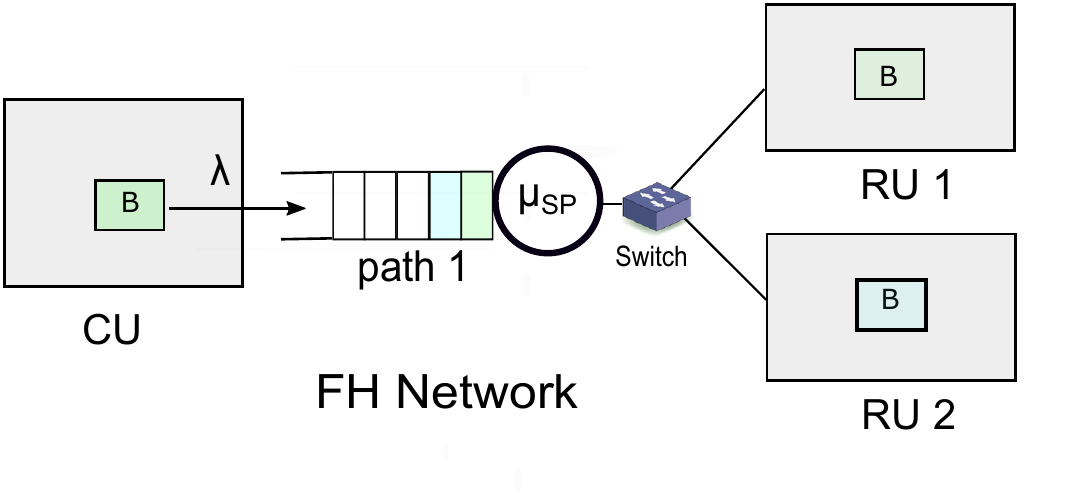}
		\caption{Single-path FH transport for downlink communication.}   \label{fig:single_path_model}
	\end{centering}
\end{figure}

\subsection{FH Technologies}
The following communication technologies and protocols are possible solutions for the deployment of FH links (see, e.g., \cite{7456186}).

\begin{itemize}

\item Dedicated fibre links: Dedicated fibre optics FH links for all RUs are attractive for their low latency and high capacity, as they can support up to $40$ Gbps per channel. However, the deployment cost generally makes it prohibitive to connect many RUs to a CU using this technology. 

\item Dedicated microwave links: Microwave FH links can offer rates from $10-100$ Mbps up to $1$ Gbps, depending on the range and weather conditions. The range limitation and sensitivity to weather events limit the scalability of the technology and its capability to sustain the traffic growth of LTE/LTE-A beyond 2017-2018 \cite{Vodafone—Mobile}.

\item Shared Coarse Wavelength Division Multiplexing (CWDM): When fibre resources are limited, it is possible to multiplex FH channel on a link using CWDM, which can provide an overall throughput as high as $100$ Gbps and a latency as low as $5~\mu$s.

\item Passive Optical Networks (PON): PON is currently used as a low-cost solution to deploy a fibre optics-based FH multi-hop network. Among all forms of PONs, Time Division Multiplexing PON is seen as a cost-effective candidate as it is able to share optical fibres and transmission equipments across multiple FH connections \cite{6886953}. Specific examples include Gigabit-PON, which provides $2.5$ Gbps downstream and $1.25$ Gbps upstream, and Gigabit Ethernet PON (G-EPON), which is being upgraded to 10G-EPON by IEEE 802.3a, offering data rate in the order of $10$ Gbps downstream and upstream. However, it does not satisfy the latency requirement of 5G, especially on the upstream.

\item Ethernet multi-hop FH: Ethernet FH multi-hop networks can help reduce cost and simplify network deployment and management by sharing the network infrastructure among multiple RUs and Cloud-RAN systems through its packet-switched operation. Another major benefit of Ethernet is its capability of flexibly scaling with the dynamic nature of data traffic. However, its use for fronthauling imposes many challenges, such as lack of synchronization, high latency and high jitter.

\end{itemize}

\subsection{Functional Split in Cloud-RAN}

In standard deployments of Cloud-RAN systems, the Common Public Radio Interface (CPRI) protocol is used to transport baseband samples between BBU and RUs. CPRI requires an end-to-end latency of around $250$ $\mu$s and the support of a high rate. As an example, four antennas covering three sectors require a $14.7$ Gbps throughput for $20$~MHz bandwidth channels. To this end, CPRI compression is commonly used in order to reduce FH data rate \cite{6772137,park2014fronthaul}. Nevertheless, given the significant increase in users' data rate and in the number of antenna in 5G, CPRI data rate can have prohibitively large values for 5G deployments.

To overcome the high data rate requirement of CPRI, alternative CU-RU functional splits have been introduced, whereby more functions are shifted into RU \cite{3gpp.38.801}. One of the key advantages of these alternative splits with less stringent FH latency constraints is that they enable the use of packet-based FH networks. Most notably, Cloud-RAN with Ethernet-based FH has received tremendous attention given that Ethernet is a widely deployed technology, it is cost-effective, and it relies on off-the-shelf standard equipment \cite{kar50278}. It also allows network monitoring and orchestration by means of virtualization and Software Defined Network; it enables sharing and convergence with Ethernet-based fixed networks; and it benefits from statistical multiplexing as a means to efficiently utilize the network capacity. On the flip side, Ethernet-based FH requires packetized communication, which imposes challenges in delivering synchronization, and meeting desired latency and jitter requirements.  Accurate phase and frequency synchronization can be attained by means of the Precision Time Protocol or via a master-slave mechanism \cite{7247590}. Furthermore, delay and jitter can potentially be reduced by using low-latency switches and applying management of path set-up.

The performance of different functional splits in the presence of Ethernet-based FH is studied in number of experimental and simulation studies \cite{6882691, 8252881}. For functional splits within the PHY, the experiments in \cite{7830261} addressed the split between wireless channel coding and wireless modulation within the PHY. For layer-2 functional split, i.e., split between PHY and MAC layers, some of the authors of this work demonstrated its feasibility for different classes of 5G traffic in \cite{7925770,8269135}. The authors in \cite{8255992} presented a flexible selection of most relevant splits, where the objective is to minimize the inter–cell interference and the FH utilization. Finally, recent advances on CPRI specification for 5G delivered enhanced CPRI, which is designed to support different functional split options \cite{ecpri}.

\section{System Model and FH Solutions}\label{system_model}

In this section, we first introduce the system model for the Cloud-RAN system under study and then detail the SP, MPD, and MPC FH transport strategies. 

\subsection{System Model}

As illustrated in Fig. 1, we consider the downlink of a Cloud-RAN system with a single CU and a number of RUs, which are connected by a multi-path FH network with $n$ pre-defined and distinct paths. Each path ends at a switch that can deliver a MAC frame to any of the RUs with negligible delay (see also Fig. 3 and Fig. 4).  To simplify the discussion, we henceforth consider the case of two RUs, but the treatment applies more generally. Each path is modelled by a queue with a single server and an exponential service time with rate $c$ bits per second. More specifically, in order to transfer a MAC frame of $B$ bits, each queue takes an amount of time that is exponentially distributed with mean $1/\mu=B/c$ seconds. Note that the queuing model abstracts the details of the multi-hop paths. Downlink traffic arrives at the CU with arrival rates and frame sizes that depend on the service type, as it will be further discussed in the next sections. As indicated in its metadata, each frame needs to be delivered to either of the RUs.


\subsection{FH Solutions} \label{FH-models}

As discussed, in order to improve reliability of FH transfer, one can in principle apply retransmission methods based on feedback, multi-path transmission with duplication, or MPD, and multi-path transmission with coding, or MPC. The improvement in reliability generally comes at the expense of latency. In the case of retransmissions, the need for feedback and for additional protocol control messages can increase the transmission time in a highly nonlinear manner \cite{6466405}, making it a non-viable solution for fronthauling. Not requiring any form of feedback, MPD can offer better latency performance, but the congestion caused by the transmission of duplicated packets over multiple interfaces can still entail unacceptable latency levels \cite{8101523}. More generally, MPC can add controlled redundancy in order to obtain a desired trade-off between reliability and diversity. In the context of packet-based FH, the most relevant coding schemes are erasure codes, which enable the recovery of a data stream, despite missing packets, as long as a sufficiently large number of encoded packets are received. We now detail the SP, MPD, and MPC FH solutions.

\subsubsection{Single Path (SP)}

As illustrated in Fig. \ref{fig:single_path_model}, with SP, the CU uses a single path to communicate to both RU$1$ and RU$2$, with the two data streams sharing the same path.


\subsubsection{Multi Path with Duplication (MPD)}

\begin{figure*}[t]
	\begin{centering}
		\includegraphics[width=0.7 \textwidth]{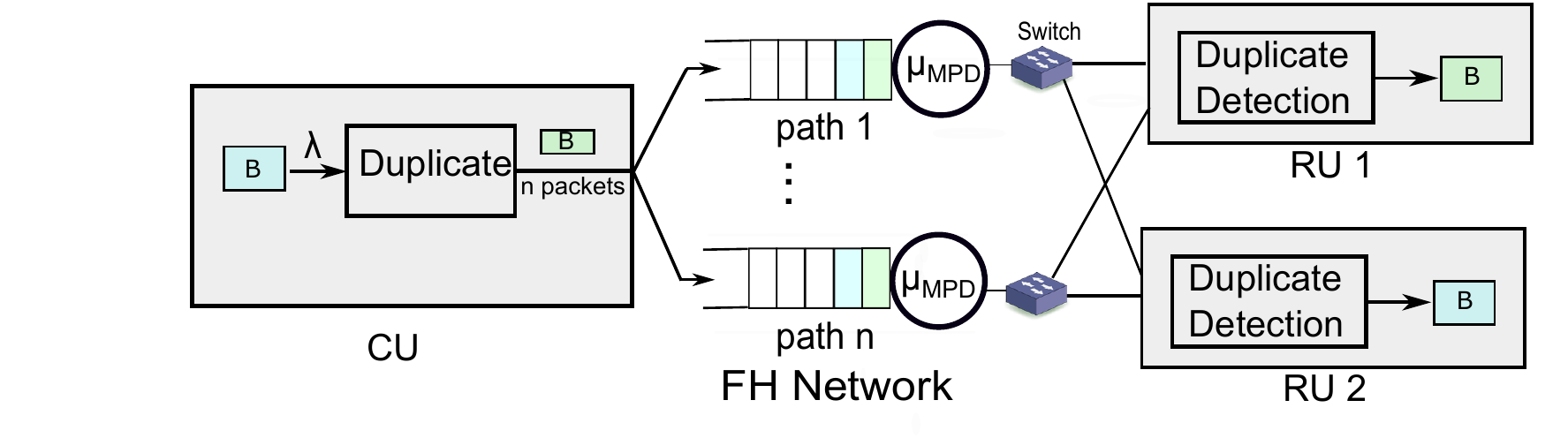}
		\caption{Multi-path FH with Duplication (MPD) for downlink communication.}   \label{fig:multi_path_model}
	\end{centering}
\end{figure*}

With MPD, as shown in Fig. \ref{fig:multi_path_model}, the CU replicates each received MAC frame, intended for either RU, on each of the $n$ FH paths. At each RU, the  \textit{duplicate detection} block detects the first successfully received frame and drops the rest.


\subsubsection{Multi Path with Coding (MPC)}\label{multi-link-coding}

\begin{figure*}[t]
	\begin{centering}
		\includegraphics[scale=1]{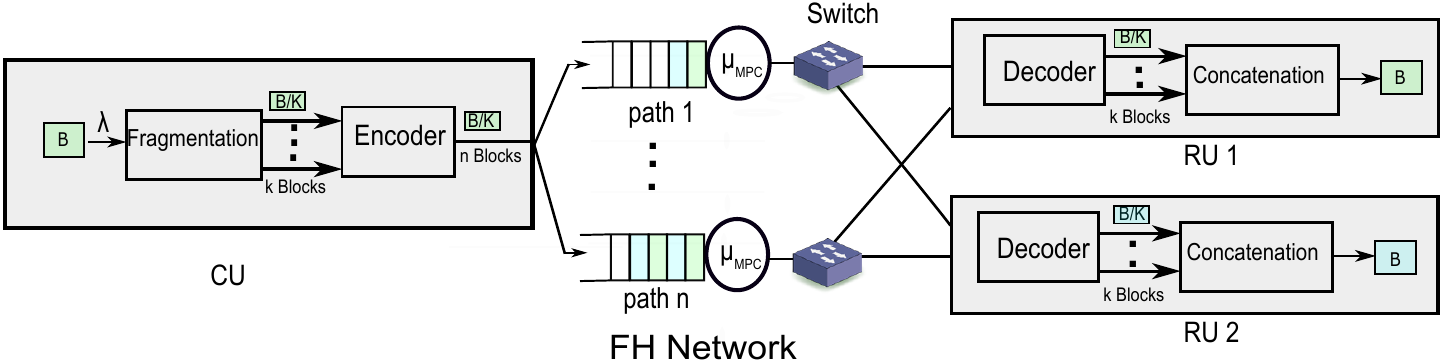}
		\caption{Multi-path FH with erasure Coding (MPC) for downlink communication.}   \label{fig:multi_path_fountain}
	\end{centering}
\end{figure*}

With MPC, an $(n,k)$ erasure code is used at the CU, with $k\leq n$. To this end, the CU carries out the following steps for each MAC frame, which may be intended for either RU:
 \begin{enumerate}
 \item Fragment each MAC frame into $k \leq n$ blocks;
 \item Encode the $k$ blocks into $n$ encoded blocks of the same size using an erasure code; 
 \item Send each block into one of the $n$ paths.
\end{enumerate}
By the properties of optimal, or Maximum Distance Separable (MDS), erasure codes, at each RU, the original MAC frame can be retrieved if any $k$ out of $n$ encoded blocks are received successfully. Note that, with $k=1$, we obtain MPD as a special case. Furthermore, setting $k=n$ yields a strategy where frames are split into disjoint segments, each sent on a different path. Overall, increasing $k$ from $1$ to $n$ gives strategies that range from MPD to frame splitting, with the former having the largest block size and the latter the smallest block size.

\section{Analysis with Single Service}\label{performance_metrics}

In this section, we analyze the performance of the C-RAN system at hand assuming a single service with random arrivals of MAC frames of size $B$ bits with exponential inter-arrival periods with average $1/\lambda$ s. We consider that each frame may be tagged independently and with equal probability as being destined to either RU. We first study the average latency required to obtain reliable FH transport and then the reliability-latency trade-off.

\subsection{Average Latency for Reliable FH Transport} \label{latency}
The analysis of the average latency is presented separately for SP, MPD, and MPC. We emphasize that the latency is defined as the time elapsed from the time packet is transmitted over the FH until the packet is successfully received.

\subsubsection{SP} \label{latency-SP}
Under the given assumptions, the average latency $T\textsubscript{SP}$ for SP can be expressed by using the standard average delay formula for M/M/1 queues, namely

\begin{equation}\label{eq:latency_SP}
T\textsubscript{SP} = \frac{1}{\mu_\textsubscript{SP} - \lambda} \end{equation} where $\mu_\textsubscript{SP}=c/B$ is the average departure rate in frames per second.

\subsubsection{MPD} \label{latency-MPD}
To analyze the performance of MPD we make the simplified assumption that, as soon as a frame is correctly received by the intended RU, all other $n-1$ copies of the same frame are deleted from the other paths. Note that this is not the case in the actual system given that it is practically difficult to remove all copies of a frame along all other paths. Therefore, the expression here provides a lower bound on the average latency. The bound is expected to be tight if the load of each path is sufficiently small. We provide some numerical evidence in the next section. 

Under the given assumption, the end-to-end system can be studied as an M/G/1 queue in which the service time is the first-order statistic of $n$ exponential variables, each with mean $1/\mu\textsubscript{MPD}=B/c$. Therefore, the average latency $T\textsubscript{MPD}$ in the original set-up can be lower bounded by using the Pollaczek-Khinchin formula \cite{6483236}
\begin{equation}\label{eq:response_time}
T\textsubscript{MPD} = E[S] + \frac{\lambda E[S\textsuperscript{2}]}{2(1-\lambda E[S])}
\end{equation} where the mean and variance of the effective service times are given as \begin{equation}\label{eq:exp_MPD}
E[S] = \frac{1}{\mu\textsubscript{MPD}}(H_{n} - H_{n-1})
\end{equation}
\begin{equation}\label{eq:var_MPD}
V[S] = \frac{(H_{n\textsuperscript{2}} - H_{(n-1)\textsuperscript{2}})}{\mu^{2}\textsubscript{MPD}}
\end{equation}respectively, with $H_{n}$ being the generalized harmonic number defined as
\begin{equation}\label{eq:harmonic}
H_{n} = \sum_{j=1}^{n}\frac{1}{j} \quad \text{and} \quad H_{n^{2}} = \sum_{j=1}^{n}\frac{1}{j\textsuperscript{2}}
\end{equation}

\subsubsection{MPC} \label{latency-MPC}
With MPC, each frame is correctly detected as soon as the first $k$ encoded blocks are received. In a manner similar to MPD and following  \cite{6483236}, we obtain a lower bound on the average delay by considering a system in which, as soon as $k$ encoded packets are received, the rest are dropped from the remaining queues. 

As a result, the end-to-end system can be studied as an M/G/1 queue, in which the service time is a random variable $S$, distributed according to the $k\textsuperscript{th}$ order statistic of the exponential distribution with mean $1/\mu\textsubscript{MPC}=B/(kc)$. Note that the service time for each queue has mean $1/\mu\textsubscript{MPC}$ that decreases with $k$ due to the smaller encoded blocks. The average latency can be again lower bounded by using the Pollaczek-Khinchin formula (\ref{eq:response_time}) with
\begin{equation}\label{eq:expectation}
E[S] = \frac{1}{\mu\textsubscript{MPC}} \sum_{i=1}^{k}\frac{1}{n-k+i} = \frac{1}{\mu\textsubscript{MPC}}(H_{n} - H_{n-k})
\end{equation} and
\begin{equation}\label{eq:expectation}
V[S] = \frac{(H_{n\textsuperscript{2}} - H_{(n-k)\textsuperscript{2}})}{\mu^{2}\textsubscript{MPC}}
\end{equation}

\subsection{Reliability-Latency Trade-Off} \label{reliability}

In this section, we provide an approximate analysis of the reliability-latency trade-off by studying the probability that correct FH transport occurs by a given deadline $t$. The approximate analysis assumes the transmission of a single isolated frame, hence ignoring queuing delays. Accordingly, for each path, we define the probability of transporting a data packet of $B$ bytes within a latency deadline $t$ on each path as
\begin{equation}\label{eq:reliability}
F(t,B) = 1- \exp(-(c/B -\lambda)t)
\end{equation} This is the probability that the service time on each queue does not exceed $t$ when the transported block is of size $B$. We can now carry out the analysis of the reliability-latency trade-off in a manner similar to \cite{8254053}, as detailed next.

\subsubsection{SP} \label{reliability-SP}

For SP, the reliability-latency curve is directly given as 
\begin{equation}\label{eq:reliability_SP}
 F\textsubscript{SP}(t) = F(t,B)
\end{equation}

\subsubsection{MPD} \label{reliability-MPD}
For MPD, a frame is correctly decoded as long as one path is successful in delivering it, which yields the reliability-latency function
\begin{equation}\label{eq:reliability_MPD}
F\textsubscript{MPD}(t) = 1 - (1 - F(t,B))^{n}
\end{equation}

\subsubsection{MPC} \label{reliability-MPC}
With MPC, delivery is correct when any $k$ of the $n$ paths deliver an encoded packet correctly, which gives the reliability-latency function
\begin{equation}\label{eq:latency_reliability}
F\textsubscript{MPC}(t) = \sum_{r=k}^{n}\binom{n}{r}F\left(t,\frac{B}{k}\right)^{r}\left(1 - F\left(t,\frac{B}{k}\right)\right)^{n-r}
\end{equation}

\section{Experiments with eMBB-URLLC Services}\label{analysis}

In this section, we develop a simulation model to validate the analysis and to account for the more realistic scenario in the context of 5G, in which both eMBB and URLLC traffics coexist on the same Cloud-RAN system. It is noted that, while Tactile Internet applications are generally considered within the class of URLLC, it is expected that a combination of different traffic classes may be needed for the delivery of a particular Tactile Internet use case. Take remote medical intervention as an example. The application may require the transmission of a high definition realtime video stream, multiple sensors' data, as well as kinaesthesia data using a haptic device \cite{7980645}. The video stream requiring high data rate can be classified as eMBB; sensors' data as massive machine-type communications; and kinaesthesia data that requires an end-to-end latency of $1$ ms as URLLC traffic \cite{7980762}. Here, we focus on the coexistence of eMBB and URLLC services. The traffic models are based on reference \cite{160060}, where URLLC and eMBB are modelled with full buffer bursty traffic FTP model 3 \cite{36814} and IP packet size of $500$ and $1500$ bytes, respectively.

To elaborate, we assume that eMBB and URLLC traffics are independent and characterized by arrival rates and packet sizes as shown in Table \ref{table:SimulationParameters}. For each traffic, frames are tagged independently and with equal probability as intended for either RU. We have $n=10$ paths, with each path having a capacity of $c=100$ Mbps.

We compare three different coexistence strategies for the eMBB and URLLC services:
\begin{itemize}
\item \emph{FH Bandwidth Orthogonal Allocation}: Each service is exclusively given a fraction of the capacity of each path;
\item \emph{FH Path Orthogonal Allocation}: Each path is allocated exclusively to either one or the other service;
\item \emph{Shared FH}: Both traffic types share all FH paths.
\end{itemize}
For all coexistence strategies, we implemented SP, MPD, or MPC in order to control the reliability-latency trade-off. Furthermore, for the orthogonal schemes based only bandwidth or path splitting, the analysis presented in the previous section applies separately to both services, whereas shared FH requires a more complex analysis that is considered to be outside the scope of this contribution.

\begin{table}[h]
\caption{System Model parameters for 5G services}    \label{table:SimulationParameters}
\begin{center}
\begin{tabular}{|p{2.5 cm}|p{1 cm}| p{1 cm}|}
\hline
Type of traffic & eMBB & URLLC \\
\hline
\hline
Packet Size (Bytes) & 1500 & 500 \\
\hline
$\lambda$ (packet/ms) & 8 & 24 \\
\hline
\end{tabular}
\end{center}
\end{table}

\subsection{Average Latency for Reliable FH Transport}\label{latency-results}

In this section, we study the average latency required for the successful delivery on the FH of a MAC frame as investigated in Sec. IV-A.

\begin{figure}
	\begin{centering}
		\includegraphics[scale=0.35]{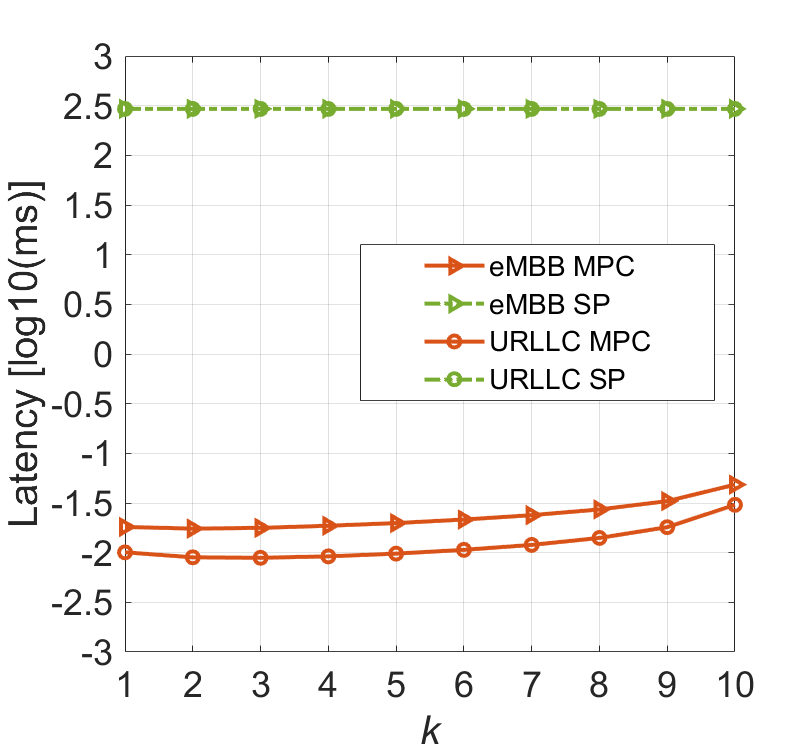}
		\caption{Average latency as a function of the frame splitting factor $k$ for SP and MPC for both eMBB and URLLC using shared FH transmission. Note that MPD corresponds to MPC with $k=1$. }   \label{fig:aver_latency_full_bw_all_k}
	\end{centering}
\end{figure}

First, in Fig. \ref{fig:aver_latency_full_bw_all_k}, we show the average latency as a function of the frame splitting factor $k$ under MPC for both eMBB and URLLC services using shared FH transmission. We also plot the performance of SP for the reference. Note that MPD corresponds to MPC with $k=1$. It is observed that MPC and MPD can drastically decrease the average latency as compared to SP for both eMBB and URLLC services. It is also seen that, under shared FH, the delay of  both services is quite close, given that the overall latency tends to be limited by queuing delays, i.e., by the time needed to traverse each shared path. Furthermore, for MPC there is an optimal value of $k$, which is around $k=2$ for eMBB and $k=3$ for URLLC. The lower optimal value of $k$ generally depends on both arrival rate and packet size of both services. This plot provides insight on how to choose $k$. For example, if the average latency for URLLC should not exceed $0.01$ ms, then the choice $k \leq 5$ would satisfy the requirement.

\begin{figure}[h]
	\begin{centering}
		\includegraphics[scale=0.35]{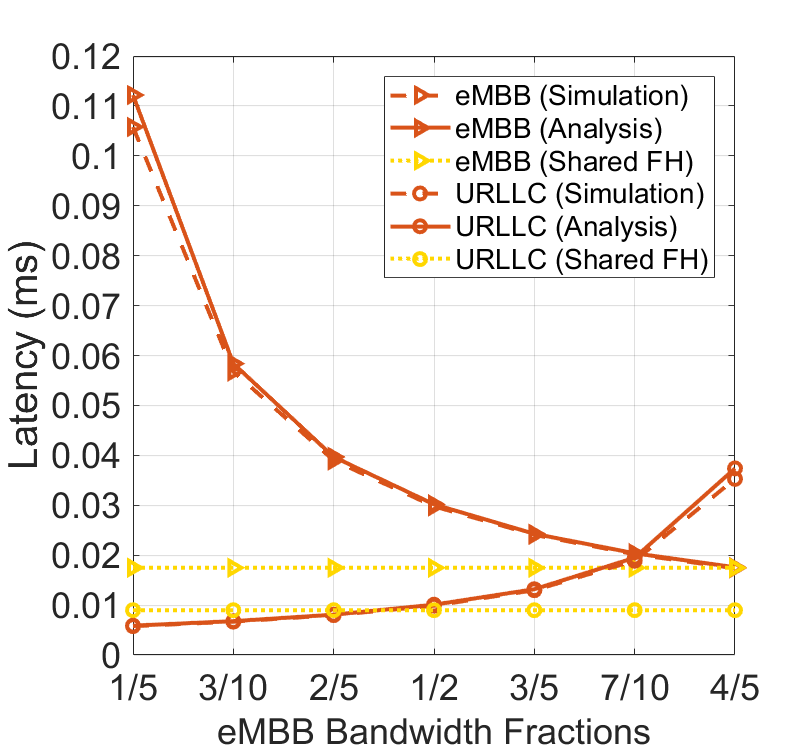}
		\caption{Average latency as a function of the eMBB bandwidth fraction for MPC with $k=2$ for both eMBB and URLLC using FH bandwidth orthogonal allocation.}   \label{fig:two_rrh_split_bw_k_2}
	\end{centering}
\end{figure}

We now consider orthogonal FH transmission schemes that can allocate a different amount of resources to eMBB and URLLC. We set $k=2$ and adopt MPC. In Fig. \ref {fig:two_rrh_split_bw_k_2}, we plot the average latency for eMBB and URLLC using orthogonal bandwidth allocation on the FH as a function of the fraction of the available path capacity, $c$, that is allocated to eMBB. We observe from the plot that, as compared to shared FH transport, orthogonal bandwidth allocation allows one to obtain a lower average latency for URLLC. Note that this is not necessarily the case for eMBB service, which is characterized by larger MAC frames. The figure also compares simulation and analysis, showing that the lower bounds derived in the previous section are tight when the load is not too high for each service.

\begin{figure}[h]
	\begin{centering}
		\includegraphics[scale=0.35]{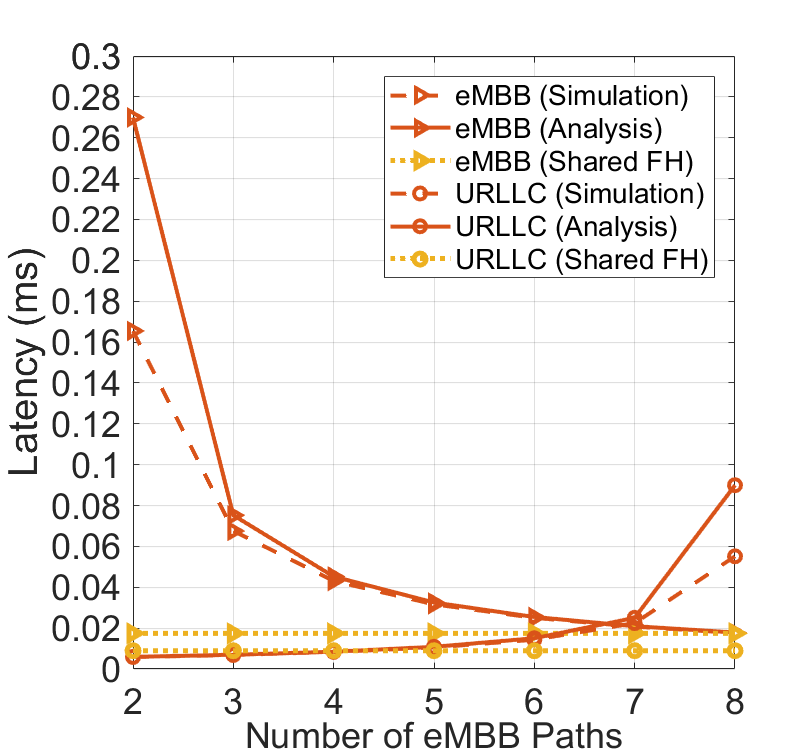}
		\caption{Average latency as a function of the number of eMBB paths for MPC with $k=2$ for both eMBB and URLLC using FH path orthogonal allocation. }   \label{fig:frac_path_equ_sim}
	\end{centering}
\end{figure}
%

\begin{figure*}[h]
	\centering
	\subfigure[eMBB. \label{fig:relia_full_bw_embb_all_stra_error}]
	{\includegraphics[scale=0.3]{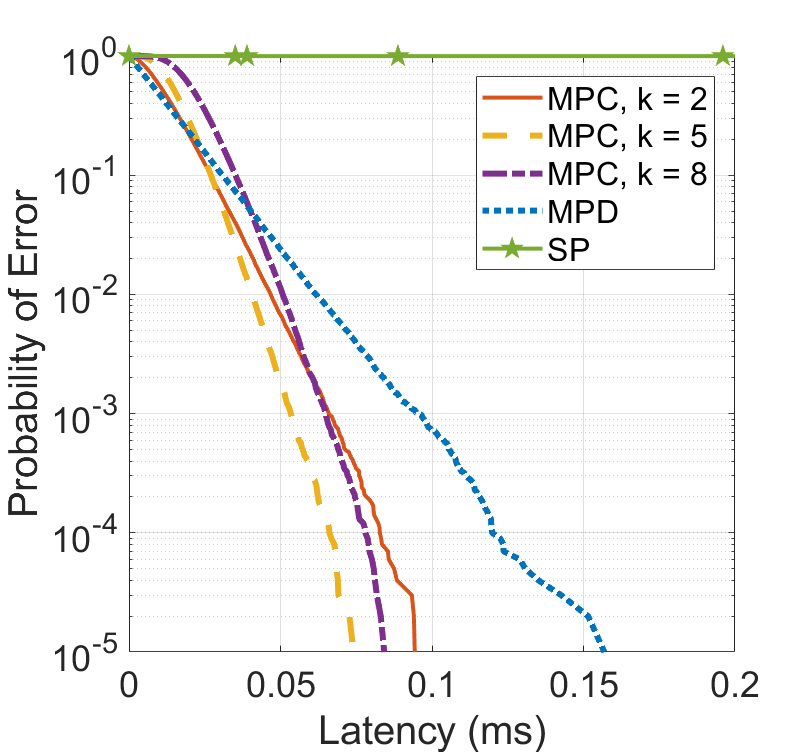}}
    \subfigure[URLLC. \label{fig:relia_full_bw_urllc_all_stra_error}]
	{\includegraphics[scale=0.3]{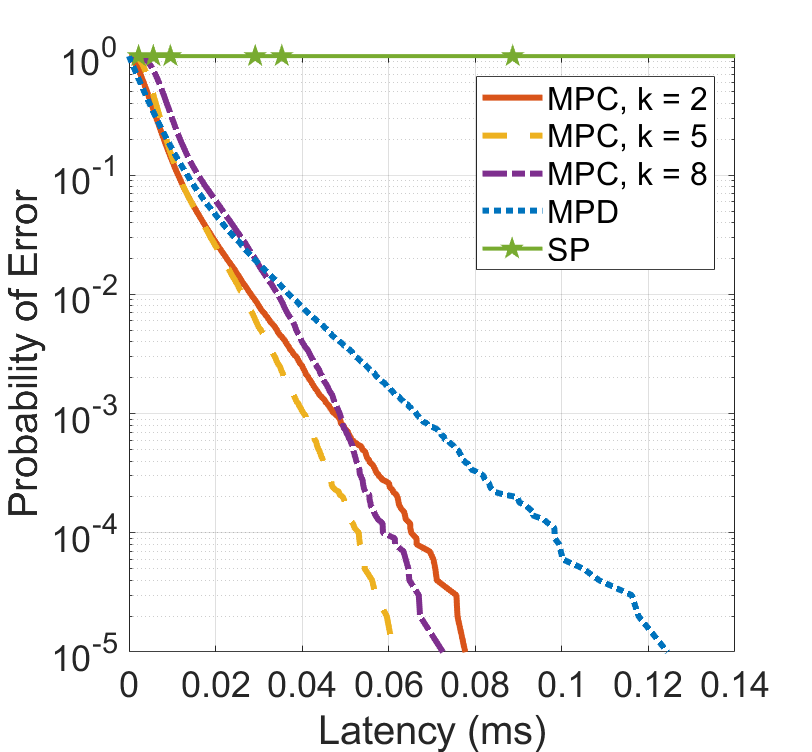}}
	\caption{Probability of error vs latency functions for SP, MPD, and MPC under shared FH transport: (a) eMBB; (b) URLLC.}
	\label{fig:relia_full_bw_all_stra_error}
\end{figure*}

\begin{figure*}[h]
	\centering
	\subfigure[eMBB.\label{fig:relia_frac_bw_embb_ratio_1_4}]
	{\includegraphics[scale=0.3]{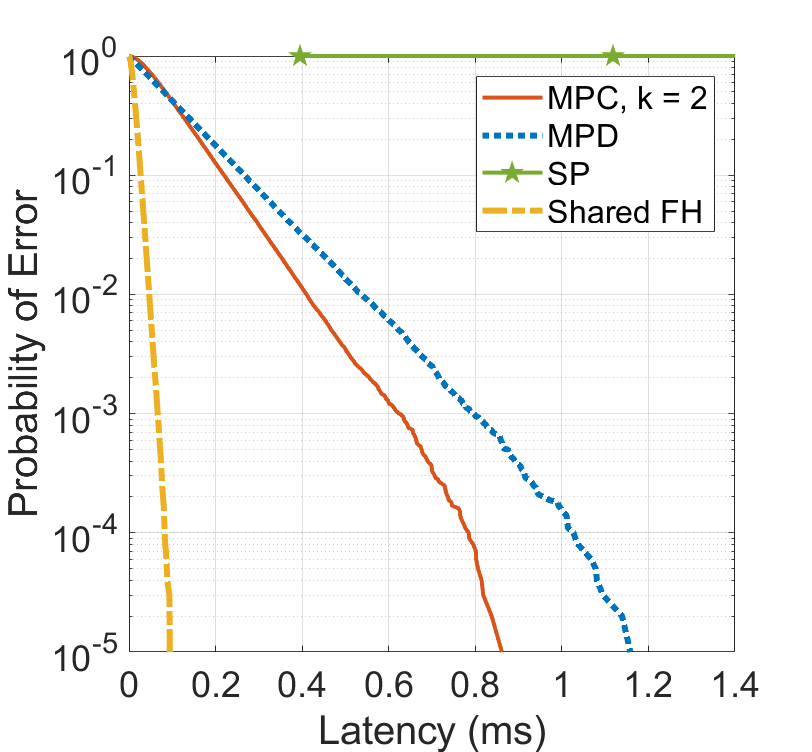}}
	\subfigure[URLLC.\label{fig:relia_frac_bw_urllc_ratio_1_4}]
	{\includegraphics[scale=0.3]{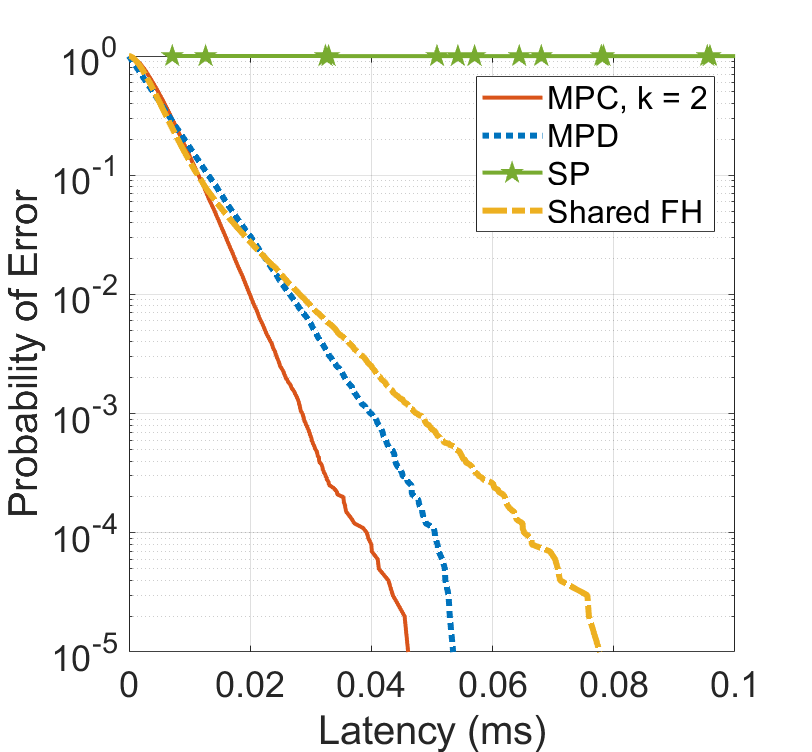}}
	\caption{Probability of error vs latency functions for SP, MPD, and MPC under orthogonal FH bandwidth split with eMBB bandwidth fraction $1/5$: (a) eMBB; (b) URLLC. }
	\label{fig:relia_bw_split_all_stra_ratio_1_4}
\end{figure*}

\begin{figure*}[h]
	\centering
	\subfigure[eMBB.\label{fig:relia_path_split_embb_ratio_1_4}]
	{\includegraphics[scale=0.3]{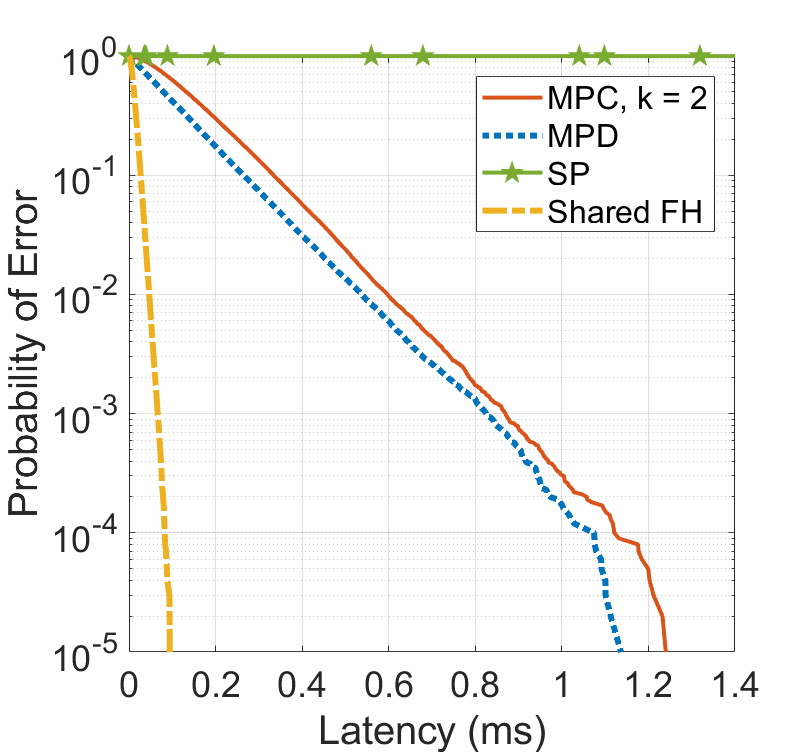}}
	\subfigure[URLLC.\label{fig:relia_path_split_urllc_ratio_1_4}]
	{\includegraphics[scale=0.3]{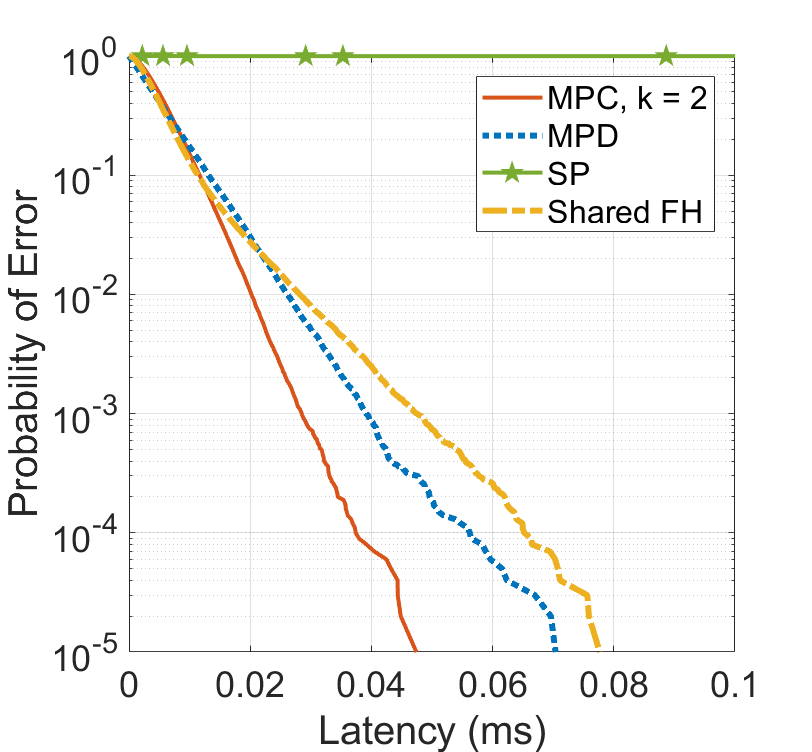}}
	\caption{Probability of error vs latency functions for SP, MPD, and MPC under orthogonal FH bandwidth split with eMBB path fraction $1/5$ ($n_e=2$ and $n_u=8$): (a) eMBB; (b) URLLC. }
	\label{fig:relia_path_split_all_stra_ratio_1_4}
\end{figure*}

To complement these results, Fig. \ref{fig:frac_path_equ_sim} shows the average latency under FH path orthogonal allocation as a function of the number of paths, $n_e$ from total $n = 10$, that are allocated to eMBB; the remainder $n_u$=$n-n_e$ are allocated to URLLC. The qualitative trend is the same as for bandwidth allocation. In particular, the average latency of URLLC can be reduced as compared to the shared FH case by means of orthogonal allocation. Furthermore, comparing path and bandwidth allocation schemes, it can be seen that bandwidth allocation can generally yield a more desirable trade-off between eMBB and URLLC performance.

\subsection{Reliability-Latency Trade-Off}

We now consider the trade-off between latency and reliability by plotting curves of reliability versus latency as defined in the previous section. In order to focus on the main regimes of interest, we plot the error probability function, which is defined as the complement of the reliability, i.e., $1-F_\mathrm{SP}(t)$ for SP and similarly for MPD and MPC. We note that the maximum requirement for error in the Tactile Internet applications is considered to correspond to the value of $10^{-5}$ as probability of error \cite{3gpp.22.261}.

As for the average latency case, we start by studying shared FH transport. In Fig. \ref{fig:relia_full_bw_all_stra_error}, we compare the reliability-latency of SP, MPD (or MPC with $k=1$), and MPC with $k=2$, $k=5$, and $k=8$ for eMBB and URLLC traffics using shared FH transmission. First, as for the average latency, we note the dramatic gains obtained by means of multi-path transport as compared to SP. Moreover, for both eMBB and URLLC, it is observed that MPC is instrumental in achieving high levels of reliability at moderate latency levels. For example in eMBB, in order to achieve a probability of error of $10^{-5}$ MPC with $k=5$ requires $0.074$ ms, while MPD entails a latency of $0.157$ ms. Furthermore, thanks to the smaller packet sizes, URLLC traffic generally attains the same level of reliability at a lower latency in the presence of shared FH.

We now consider the performance under orthogonal resource allocation. Fig. \ref{fig:relia_bw_split_all_stra_ratio_1_4} shows the probability of error as a function of the latency under bandwidth allocation with one fifth of the bandwidth allocated to eMBB. MPC with $k=2$ can achieve a latency reduction of $0.3$ ms and $0.0075$ ms in eMBB and URLLC, respectively, with respect to MPD at the error probability of $10^{-5}$. Furthermore, a larger bandwidth allocation to URLLC can significantly enhance the reliability of URLLC traffic at the cost of a larger latency for the eMBB service.

Finally, Fig. \ref{fig:relia_path_split_all_stra_ratio_1_4} considers orthogonal path allocation with one fifth of the paths allocated to eMBB, i.e. $n_e=2$ and $n_u=8$. For URLLC, we can see that MPC can reduce latency by $0.023$ ms as compared to MPD at the error probability of $10^{-5}$. Moreover, the probability of error obtained with path split is improved by approximately $60\%$ as compared to that obtained by non-orthogonal sharing of FH resources. Nevertheless, bandwidth allocation is seen to provide a better trade-off between URLLC and eMBB performance.

\section{Concluding Remarks}\label{conclusion}
In this paper, we have studied the problem of ensuring low-latency and high-reliability in a Cloud-RAN system with multi-path Ethernet-based FH network. The proposed solution is based on erasure coding and multi-path transmission on the FH network. With this approach, the CU splits the original MAC frame into smaller blocks, encodes them into a larger number of encoded blocks, and then transmits them over the multiple paths. The solution is analyzed and compared with conventional single-path FH transport and multi-path methods based on duplication. The performance is evaluated in terms of average latency for reliable delivery and of the reliability-latency trade-off. The results consider the coexistence of URLLC and eMBB traffic on the FH under both orthogonal and non-orthogonal FH resource allocation modes. As a general conclusion, MPC can achieve a low error probability of $10^{-5}$ at lower latency than MPD by means of orthogonal as well as non-orthogonal shared FH. Furthermore, in the presence of eMBB-URLLC coexistence, we showed that, by adequately managing the FH resources via orthogonal allocation transmission, the average latency and the error probability can be affectively reduced as compared to shared FH transport. These results can serve as a valuable guidance on how to effectively deploy multi-path FH resources for the implementation of Tactile Internet Applications.

\section*{Acknowledgement}
\begin{small}
This work has been mainly supported by The Engineering and Physical Sciences Research Council (EPSRC) industrial Cooperative Awards in Science \& Technology (iCASE) award and by BT. Additional support to this research is received form European Union’s Horizon 2020 Research and Innovation Programme (H2020) 5GCAR and the European Research Council (ERC) under the Grant Agreement No. 725731.
\end{small}

\bibliographystyle{IEEEtr}
\bibliography{Bibliography-TM}

\begin{IEEEbiography}
   [{\includegraphics[width=1in,height=1.25in,clip,keepaspectratio]{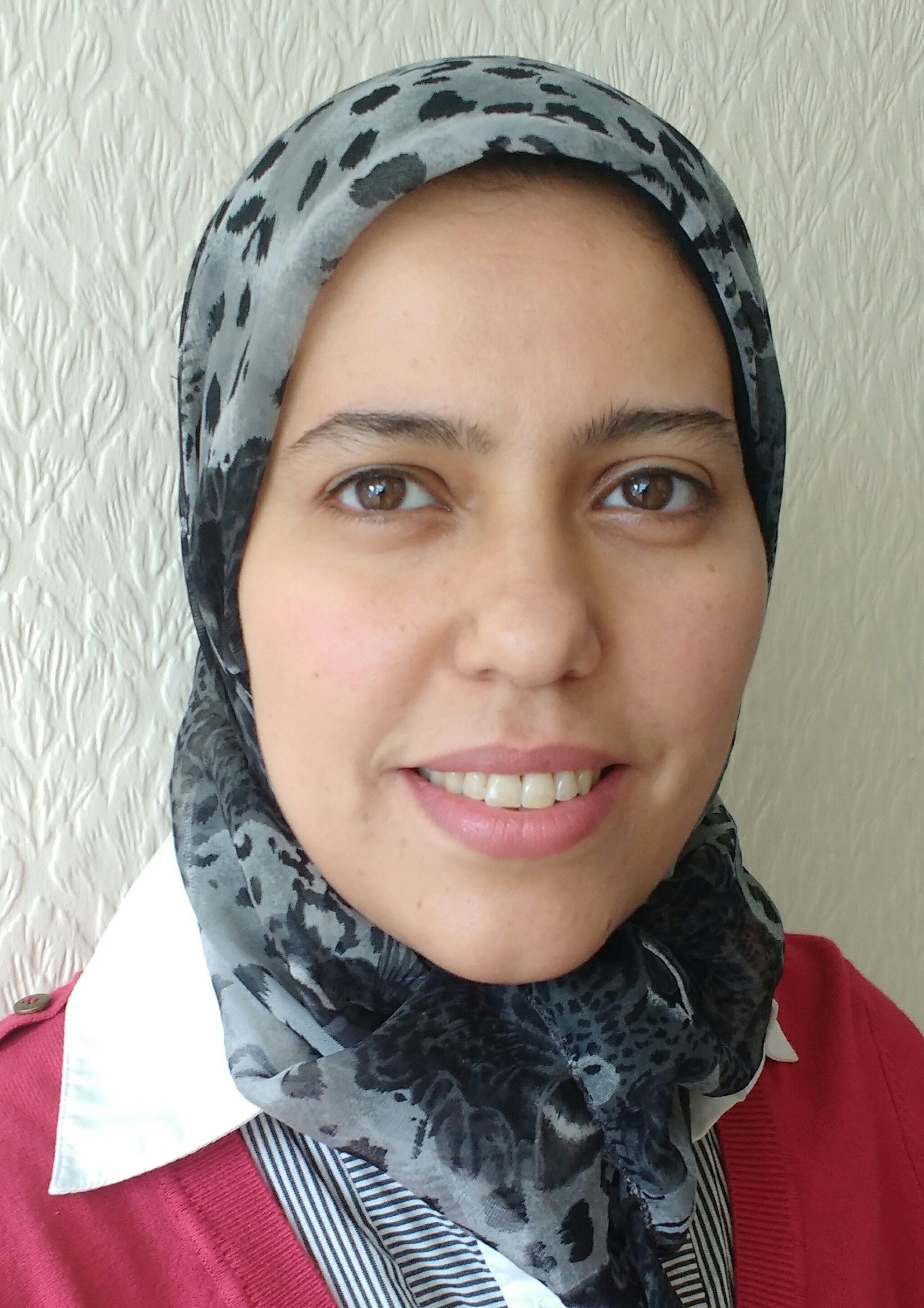}}]
  {Ghizlane Mountaser} has a BSc degree in Mobile Communications and a master degree in Mobile, Personal and Satellite Communications. She is currently a PhD student at the Centre for Telecommunications Research of King's College London, focusing on flexible functional split in 5G Cloud-RAN. Prior to this, she worked as a protocol stack software engineer in the Telecommunications industry.
\end{IEEEbiography}

\begin{IEEEbiography}
[{\includegraphics[width=1in,height=1.25in,clip,keepaspectratio]{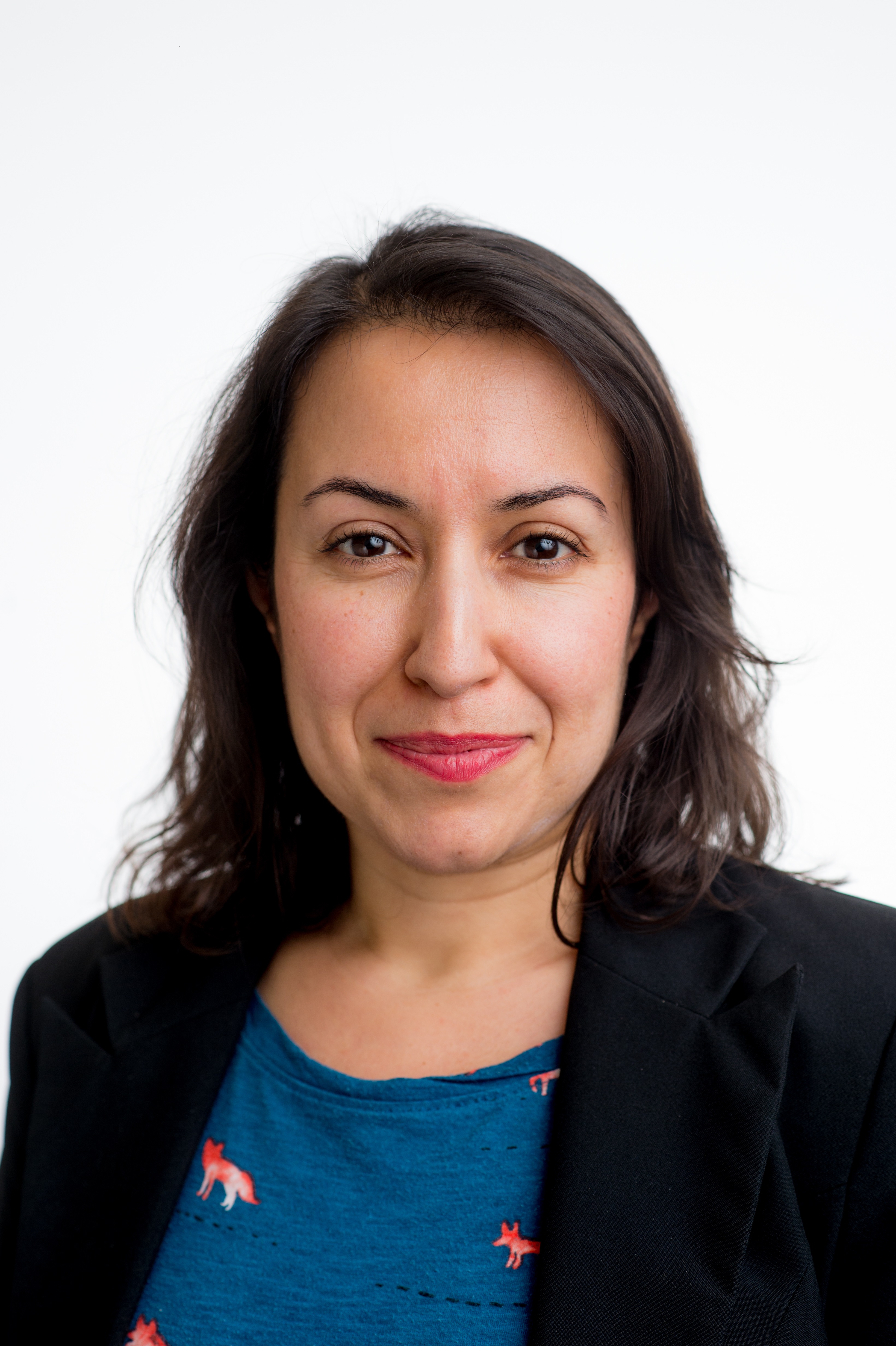}}]
  {Toktam Mahmoodi} received the B.Sc. degree in electrical engineering from the Sharif University of Technology, Iran, and the Ph.D. degree in telecommunications from King’s College London, U.K. She was a Visiting Research Scientist with F5 Networks, San Jose, CA, in 2013, a Post-Doctoral Research Associate with the ISN Research Group, Electrical and Electronic Engineering Department, Imperial College from 2010 to 2011, and a Mobile VCE Researcher from 2006 to 2009. She has also worked in mobile and personal communications industry from 2002 to 2006, and in an Research and Development team on developing DECT standard for WLL applications. She has contributed to, and led number of FP7, H2020 and EPSRC funded projects, advancing mobile and wireless communication networks. Toktam is currently with the academic faculty of Centre for Telecommunications Research at the Department of Informatics, King’s College London. Her research interests include 5G communications, network virtualization, and low latency networking.
\end{IEEEbiography}

\begin{IEEEbiography}
[{\includegraphics[width=1in,height=1.25in,clip,keepaspectratio]{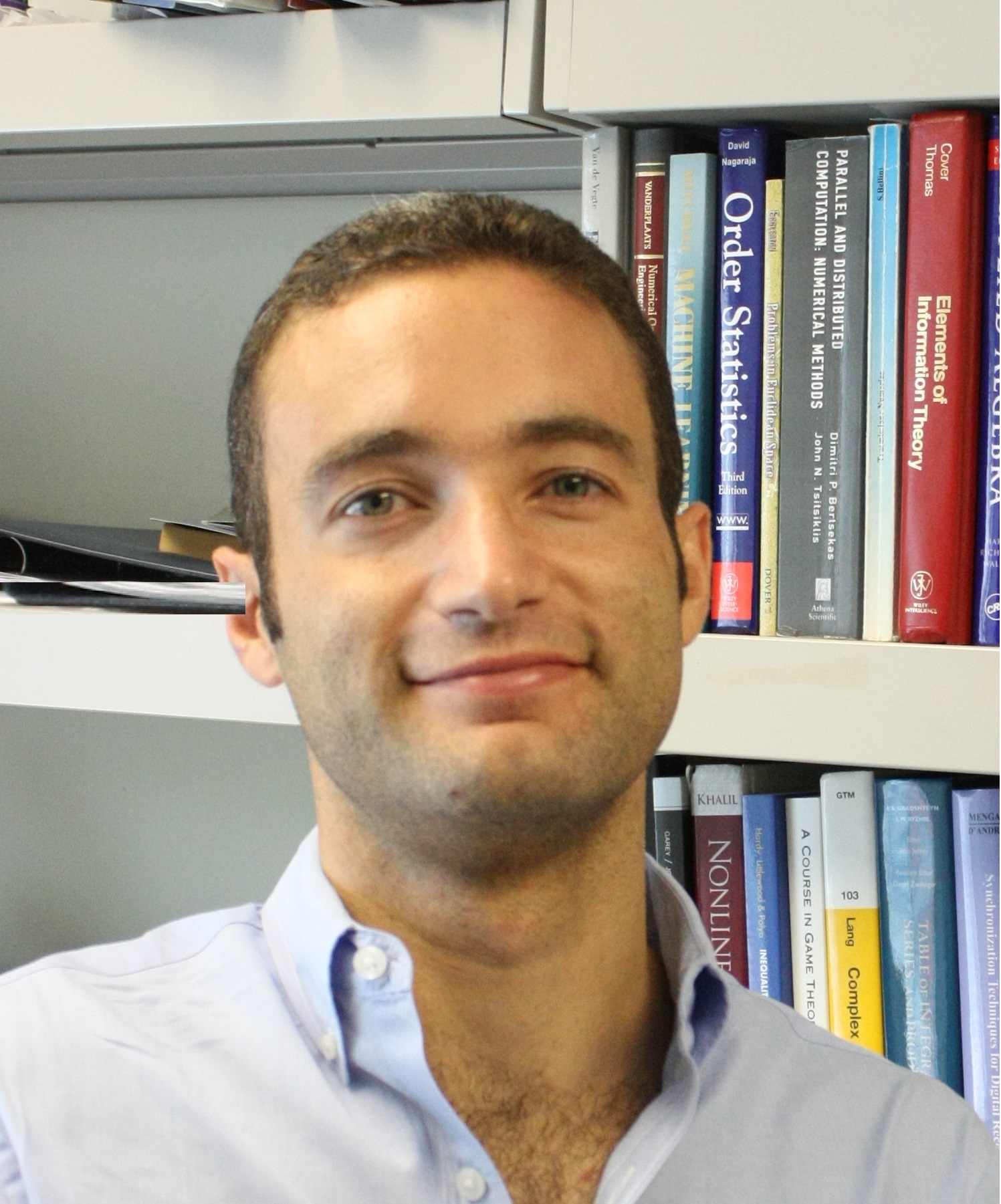}}]
  {Osvaldo Simeone} is a Professor of Information Engineering with the Centre for Telecommunications Research at the Department of Informatics of King's College London. He received an M.Sc. degree (with honors) and a Ph.D. degree in information engineering from Politecnico di Milano, Milan, Italy, in 2001 and 2005, respectively. From 2006 to 2017, he was a faculty member of the Electrical and Computer Engineering (ECE) Department at New Jersey Institute of Technology (NJIT), where he was affiliated with the Center for Wireless Information Processing (CWiP). His research interests include wireless communications, information theory, optimization and machine learning. Dr Simeone is a co-recipient of the 2017 JCN Best Paper Award, the 2015 IEEE Communication Society Best Tutorial Paper Award and of the Best Paper Awards of IEEE SPAWC 2007 and IEEE WRECOM 2007. He was awarded a Consolidator grant by the European Research Council (ERC) in 2016. His research has been supported by the U.S. NSF, the ERC, the Vienna Science and Technology Fund, as well by a number of industrial collaborations. He currently serves in the editorial board of the IEEE Signal Processing Magazine, and he is a Distinguished Lecturer of the IEEE Information Theory Society. Dr Simeone is a co-author of two monographs, an edited book published by Cambridge University Press, and more than one hundred research journal papers. He is a Fellow of the IET and of the IEEE and of the IET.
\end{IEEEbiography}

\end{document}